\documentclass[aps,prb,preprint,groupedaddress]{revtex4}

\usepackage{graphicx}
\newcommand{\be}{\begin{equation}}
\newcommand{\ee}{\end{equation}}

\begin{document}
\title{Dimensional crossover and quantum effects of gases adsorbed on
nanotube bundles}
\author{M. Mercedes Calbi and Milton W. Cole}
\affiliation{Physics Department, Pennsylvania State University, University Park,\\
Pennsylvania 16802}
\date{\today }

\begin{abstract}

Adsorption properties of several gases (Ne, CH$_4$, Ar, Xe) on the
external surface of a carbon nanotube bundle are investigated. Calculations are performed at low coverage and variable temperature, and for some fixed temperatures as a function of coverage. Within 
a simple model (in the limit of very low coverage) we are able to study the evolution of the film's thermal 
properties from those of a one dimensional (1D) fluid to those of a 2D film. In addition, grand canonical 
Monte Carlo simulations 
are performed in order to identify a second layer groove phase, which occurs
once a monolayer of atoms covers the external surface. We derive from
the simulations the isosteric heat, compresibility and specific heat as a function of coverage.
We evaluate alternative models in order to derive quantum corrections to the classical results. We compare our
findings with those of recent adsorption experiments.

\end{abstract}

\maketitle

\section{Introduction}

The nature of gas adsorption within and outside of bundles of carbon
nanotubes is a burgeoning field, which promises to reveal novel phases of
matter \cite{rmp,will,wang,boro,bon,dil,bob,aldo,aldo2,aldosg,aldosg2,yates,oscar1,oscar,kaneko,kan&bill}. Several recent experiments have explored the behavior of gases
adsorbed on the outside of bundles, or ropes, consisting of many nanotubes
having nearly parallel orientation. Our group has studied this system with
classical computer simulation (in the case of Ne, Ar, Kr and Xe) and model
ground state calculations purporting to describe this external surface
adsorption of classical, or nearly classical, gases \cite{exsurf1,exsurf2}. Other studies, both
experimental and theoretical, have been undertaken of the behavior of
quantum gases, such as hydrogen or helium, in quasi-one dimensional (1D) models of this environment \cite{boro,bon,dil,bob}.

Figure 1 shows the potential energy $V({\bf r})$ of a methane molecule on the external surface of a nanotube bundle,
computed (as discussed in Section IV) by summing empirical Lennard-Jones 
pair
interactions between the molecule and the C atoms comprising the nanotube
(which are actually smeared out to form ``continuum carbon'' on a
cylindrical surface) \cite{george}. The characteristic features of V are a small region nestled between tubes, called a ``groove'', in which the potential is extremely
attractive, and a more extended region where the potential varies only
slowly along the surface. It is not surprising that the adsorption at a given temperature
(computed
in Section IV) exhibits a 1D fluid regime, confined to the groove, and a 2D
monolayer regime as the number of adsorbed particles increases. The 
evolution of this behavior as a function of gas
pressure (P) and temperature (T) is qualitatively the same as that found in
our simulations of the other gases \cite{exsurf1,exsurf2}.

At least some of the extant experimental data \cite{aldo,aldo2,aldosg,aldosg2,yates,oscar1,oscar,kaneko} are in qualitative, or even quantitative, agreement with calculations based on such simple
model assumptions about the geometry (nanotubes are uniform, identical and
parallel to one another) and interactions (pairwise additive) \cite{milen}. As an example, we state in Table I the ratio between the binding energies in the groove and on graphite in the low density limit, obtained from our calculations and two different experimental groups. Thereby
encouraged, we proceed in the present paper to address several aspects of
this adsorption problem that were either ignored in our previous studies or
received only cursory attention there. One particular goal is to better
understand how the film's thermal properties evolve from those of an
essentially 1D fluid to those of a monolayer (and eventually bilayer) film
adsorbed on this bundle's surface. An aspect of this investigation is to
identify and characterize a ``second layer groove phase'', for which
simulations and experiments have provided some evidence \cite{aldosg,aldosg2,exsurf1,exsurf2}. A second principal
goal of the present study is to understand how quantum effects alter the
thermodynamic quantities of interest. These effects are particularly important for H$_{2}$, He, Ne and CH$_{4}$ films in the present geometry \cite{wang}. We emphasize that the potentials we use, like all semiempirical potentials, are uncertain to some degree. As an example, we may compare the well depths computed with the present parameters and those resulting from a recent selection of ``best values''\cite{bruch}. The well depth from a single graphene sheet is given by $D_{graphene}=(6\pi/5)\epsilon_{gc}\sigma_{gc}^2\theta_c$, where $\epsilon_{gc}$ and $\sigma_{gc}$ are the gas-carbon Lennard-Jones parameters and $\theta_c$ is the surface density of carbon atoms. The ratio of this depth computed with the ``best'' parameters to that obtained from our parameters is 0.8, 0.8, 1.0, 1.0, 1.1, 1.1 for the sequence He, Ne, Ar, Kr, Xe and H$_2$, respectively.

In Section II, we consider the behavior of a film at very low coverage, in
which case the adsorption may be considered as a single particle problem;
this corresponds to the so-called ``Henry's law regime'' of adsorption \cite{henry}, in
which the coverage N is proportional to the pressure, as specified by the
Henry's law constant K$_{H}=\lim_{N \rightarrow 0}(N/P)$. This ``constant'' typically exhibits an Arrhenius dependence on T, with
characteristic activation energy of order the heat of adsorption. As is
known from analogous treatments of adsorption on planar surfaces, the T
dependence of K$_{H}$ is a sensitive function of the potential provided by
the substrate, a subject of obvious interest \cite{bruch}. We compute in this low coverage regime the specific heat (per particle) $C/N$, where $C$ is
the heat capacity, and the isosteric heat

\begin{equation}
q_{st}=-\left( \frac{\partial \ln P}{\partial \beta }\right) _{N} 
\end{equation}

Here $\beta ^{-1}=k_{B}T$. Further, in Section II, we develop a simple
``crossover potential'' model, aimed at describing the film's behavior in
the regime of crossover from 1D to 2D (here, a consequence of increasing T).
The model works well in characterizing the thermal properties of a low
coverage classical film, as demonstrated by consistency between the results
from the model and those from the ``exact'' simulations. This agreement
provides some justification for using the model to perform quantum
calculations for this low coverage regime. The primary result of this
calculation is a determination of the quantum correction to the thermal
properties computed in the simulations.

Section III develops a phonon model for the high density 1D fluid regime of
adsorption within the groove. The treatment is a straightforward extension
of the familiar Debye model used to treat bulk solids. There is an
underlying premise- that phonons are, indeed, the relevant thermal
excitations of a 1D fluid (since there is no crystallization), from which
the thermal properties may be calculated. The reliability of this ansatz is
discussed and a heuristic argument is provided to justify its use here.

Section IV presents new classical simulation results for $q_{st}$, $C$ and the
compressibility $\left( \partial N/\partial \mu \right) _{T}$, where $\mu $
is the chemical potential. Methane adsorption isotherms are presented for several temperatures. A key question of interest is whether the
hypothetical second layer groove phase exists for all of the adsorbates we
have considered, or just for Ne, as we originally conjectured. In addressing
this issue, we make contact with a growing body of experimental data \cite{aldosg,aldosg2}. Section V summarizes our results and discusses other issues, such as the
potential role of heterogeneity, which is certainly present, but is ignored
in virtually all calculations to date.

\section{Low density regime}

At very low density, one may neglect both interactions and quantum
statistics in evaluating the film's properties. This means that the physical
behavior of the system reflects the dynamics of single particles in the
external field of the surface. Let $g(E)$ be the surface single particle
density of states, which may be derived, in principle, by solving the
Schrodinger equation in this potential field. Then, the total number of
adsorbed particles at T and $\mu $ satisfies

\begin{equation}
N=e^{\beta \mu }\int dE\, g(E)e^{-\beta E}=e^{\beta \mu }Q(\beta ) 
\end{equation}

Here $Q(\beta)$ is the single particle partition function. For an 
ideal spinless \cite{spin} 3D
gas, assumed to be in equilibrium with the film, $e^{\beta \mu }=\beta
\lambda _{T}^{3}P$, where $\lambda _T $ is the de Broglie thermal 
wavelength ($\lambda _{T}^{2}=2\pi \beta \hbar ^{2}/m$)\ . 
By differentiating Eq. 2, at
fixed N, we obtain the quantum expression for the isosteric heat at low density:

\begin{equation}
q_{st}=\frac{5}{2\beta }-\langle E\rangle =\frac{5}{2\beta }+\left( \frac{\partial
\ln Q}{\partial \beta }\right) _{N,A} 
\end{equation}

$\langle E\rangle $ is the quantum-statistical average of the energy per particle. The
quantum value of the Henry's law coefficient satisfies

\be
K_{H}=\beta \lambda _{T}^{3}Q(\beta ) 
\ee

The quantum specific heat per particle (expressed in units of Boltzmann's
constant) at fixed surface area $A$ satisfies

\be
\frac{C}{Nk_B}=\beta ^{2}\left( \frac{\partial ^{2}\ln Q}{\partial \beta ^{2}}\right) _{N,A} 
\ee

\noindent and from Eq. 3,

\be
\frac{1}{k_B}\frac{dq_{st}}{dT}=\frac{5}{2}-\frac{C}{Nk_B}
\ee 

Classically, one employs an alternative formulation of the behavior of an
ideal gas in the external field of the substrate. The Henry's law constant
is then

\be
K^{cl}_{H}=\beta \int d{\bf r}\, e^{-\beta V({\bf r})}=\beta Z(\beta ) 
\ee

$Z(\beta )$ is a classical single particle configuration integral and
the integration domain is a (somewhat arbitrarily defined) volume near
the substrate.  The classical isosteric heat differs from the quantum
expression because the quantum mean energy is replaced by the
classical energy, $\frac{3}{2\beta} +\langle V\rangle _{cl}$:

\be
q_{cl}=\frac{1}{\beta }-\langle V\rangle _{cl}=\frac{1}{\beta }+\frac{%
d \ln Z}{d \beta } 
\ee

The classical specific heat is given by

\be
\frac{C_{cl}}{Nk_{B}}=\frac{3}{2}+\beta ^2 \frac{\partial ^{2}\ln Z}{\partial
\beta ^{2}} 
\ee

For reference, we present in the Appendix analytical results for the limiting
case of motion in 1D with a quadratic transverse confining potential

\be
V(r)=V_{0}+\alpha \frac{r^{2}}{2} 
\ee

\noindent where $r^{2}=x^{2}+y^{2}$ and the $z$ axis is the usual cylindrical symmetry
axis, coincident with the groove site.

The full line in Fig.2 depicts the classical specific heat in 
the limit of zero coverage
computed from the potential energy shown in Fig.1. One
observes that the behavior crosses over from the 1D value (5/2, as shown in the Appendix) at low T to a value near the 2D value (2, since $y$ motion is included) at high T. However, there appears a dramatic peak at
intermediate T $\approx$ 200 K. This behavior can be explained qualitatively, as follows. 

We invoke a simple ``crossover model'' of the potential which divides configuration space into two regions, giving additive contributions to the partition function:

\be
Z(\beta)=\int_{groove} d^2r \, e^{-\beta (V_g+ \frac{1}{2}
\alpha r^2)}+ L_s \int_{mono} dy\, e^{-\beta (V_m+ \frac{1}{2}
k_m y^2)}   
\ee

The first integral is a 2D integral over the groove region, where we assume the potential energy to be of harmonic type. The second integral evaluates the contribution from the monolayer region over the external surface of the tubes, assuming a harmonic potential along the direction perpendicular
to the surface. $L_s$ is the length of this region (a fraction of $2\pi R$, $R$ being the radius of the tubes) over which the monolayer states can extend. We estimate reasonable values for each constant in the model (radius of the groove region $\approx 1$ \AA, $V_g = -2028$ K, $\alpha = 4800$ K/\AA$^2$, $V_m = -1050$ K, $k_m = 6000$ K/\AA$^2$, $L_s = 18$ \AA) from the full potential of Fig. 1.
Using this model partition function, we obtain the results shown in Fig. 2 as a thin line. The position and shape of the peak in Fig. 2 agree well with the ``exact'' results. This implies that the model has the key physical ingredient, which is the crossover from 1D to 2D regimes. We observe that the peak temperature ($\approx$ 220 K) is significantly less than the energy difference between monolayer and groove potentials ($V_m - V_g \approx 1000$ K).

However, we note qualitative differences between the ``exact'' and model curves at both low and high T. These differences come from the deviation of the full potential from the model's assumed harmonic shape. First, we consider the linear region of variation of C at
low T. This behavior may be derived from perturbation theory, assuming that
there is a ``small'' quartic term in the potential, $\delta V$, which may be expressed
in terms of appropriate coefficients $a_{i}$:

\be
\delta V(x,y)=a_{1}x^{2}y^{2}+a_{2}(x^{4}+y^{4}) 
\ee

The classical energy shift $\delta E$ is just the expectation value of the
perturbation (evaluated with the harmonic potential, Eq.10); this leads to \cite{qanah} 
$\delta E=(a_{1}+6a_{2})/(\alpha \beta )^{2}$. This quartic correction to $E$ gives rise to terms proportional to T in the specific heat that are observed in Fig. 2 at low T, i.e. the regime ($T<100$ K) where the quartic 
terms are adequately described by perturbation
theory. At very high T, the deviation of $C/(Nk_B)$ from the 2D value (2) occurs because of the large $y$ behavior of the full potential, which lies below the harmonic approximation.

The simple crossover model potential is thus seen to yield a specific heat
which agrees qualitatively with the numerical results (without fitting the
parameters). Hence, we conclude that the model contains the essential
ingredients possessed by the ``true'' potential. We now use the simple model
potential in another way to deduce the quantum expression for the specific
heat and other thermal properties at very low N. To do the quantum
calculation accurately requires a determination of the density of states in
this inhomogeneous environment. Our alternative method involves a further set of
approximations, which obviate solving the full 3D Schrodinger equation (SE). 
An exact solution is not necessary, in our opinion, given the uncertainty
in the potential V({\bf r}) and the manifestly satisfactory character of the
simple crossover model potential. To obtain the energy spectrum, we solve
the SE with an adiabatic approximation, which has been shown to predict band
structures for He atoms on graphite \cite{adiab}. In this approximation, we first solve
the SE for adatom motion perpendicular to the surface at each point ($x$) on
the surface. The eigenvalue for this z motion becomes an effective potential
energy V$_{eff}$($x$). As in the conventional treatment of diatomic molecules
(focusing on the electronic ground state), we assume that the lowest state
for the adatom's $z$ motion is the only energy needed for the ``slower''
degree of freedom, i.e. motion along the surface. Figure 3 shows the
resulting effective potential as well as a fit to this potential of the
following form:

\be
U(x)=V_{m}+\frac{V_{0}}{\cosh ^{2}(x/a)} 
\ee

The reason for this choice of model potential is that the shape is
appropriate, as seen in the figure, and the spectrum is known for this
functional form, as given by \cite{qmbook}

\be
\epsilon _{n}= -\frac{\hbar ^2}{2m a^2}\left[ \frac{1}{2}\sqrt{\frac{8m V_0 a^2}{\hbar ^2}+1}-\left( n+\frac{1}{2}\right) \right] ^2\hspace{1cm}; \,\, n = 0,1,2...
\ee

The largest value that $n$ can take depends on the parameters of the model potential. For example, we obtain 21 discrete levels for CH$_4$, 11 for Ne and 3 for H$_2$. With this spectrum, we can factorize $Q(\beta )$ in the following way:

\be
Q(\beta )=\frac{L}{\lambda _{T}}\left[ Q_{-}(\beta )+Q_{+}(\beta )\right]
Q_{y}(\beta ) 
\ee

\noindent where $L/\lambda _{T}$ is the sum of the free particle states along the $z$
direction, $Q_{y}(\beta )$ comes from the contribution of the discrete
states in a potential well along the direction perpendicular to the surface
(here assumed to be a harmonic oscillator's spectrum):

\be
Q_{y}(\beta )=\mathrel{\mathop{\sum }\limits_{n}}
e^{-n\beta \hbar \omega _{y}}=\frac{1}{1-e^{-\beta \hbar \omega}} 
\ee

$\left[ Q_{-}(\beta )+Q_{+}(\beta )\right]$ is the partition function of all
the states in the effective potential along the $x$ direction. The first term (%
$Q_{-}$) comes from the discrete states in $V_{eff}(x)$:

\be
Q_{-}(\beta )=%
\mathrel{\mathop{\sum }\limits_{n}}%
e^{-\beta \epsilon _{n}} 
\ee

The contribution $Q_{+}$ from the continuum density of states is modeled by
scaling the continuum spectrum of a uniform system in a constant potential of depth $%
V_{m}$ by the length of a unit cell $d$ (here equal to the horizontal
separation between nanotubes):

\be
Q_{+}(\beta )=e^{-\beta V_{m}}\frac{d}{\lambda _{T}} 
\ee

With these ingredients, the thermodynamic properties can be computed and
compared with the corresponding classical quantities, obtained above from
the original potential. The dashed line in Fig. 2 show the quantum result for C. In this case, the low T behavior corresponds to a quantum 1D regime characterized by the excitation of only the low lying states 
in $V_{eff}$. Note that the zero temperature limit is 0.5, as expected for
 a quantum 1D system. Deviation from that limit occurs when T comes within an order of magnitude of the transverse excitation energy in the groove ($\approx 90$ K for CH$_4$). This behavior differs markedly from the classical result in Fig. 2. Fig. 4 shows the corresponding results for the isosteric heat. At zero T, the classical result (full and thin lines) corresponds to the minimum of the potential in the groove while the quantum result is the ground state energy $\varepsilon_0$ in that potential. Another important distinction at low temperature is that quantum results imply an increase of the isosteric heat with T, whereas classically a decrease is expected. We observe that $q_{st}=-\varepsilon_0+2k_BT$ when $T<50$ K and it reaches its maximum value when $C/(Nk_B)=5/2$ (Eq. 6) at $T\approx 100$ K. As shown in the Appendix, at low T the difference between quantum and classical heats is $\delta q_{st} = 2/\beta - \hbar \omega$, where $\hbar \omega$ is the zero point energy in the groove. At high T, the temperature dependence of $q_{st}$ is given by $k_BT/2$, which corresponds to 2D motion confined by a transverse harmonic potential. Figure 5 (a) and (b) display the results for C and $q_{st}$ in the case of Ne. We notice that the crossover occurs at a much lower T ($\approx 100$ K) for Ne than for CH$_4$.

In closing this section concerning low density, we note that the first order correction
due to quantum statistics can be determined in a straightforward way,
permitting an assessment of its importance in the analysis. To do this, one
expands the exact equation relating N and $\mu $, assuming that the fugacity $e^{\beta \mu}$
is small; this is the usual way to develop a quantum virial expansion for
translationally invariant systems. The result in the present case is

\be
N=Q(\beta )e^{\beta \mu }\left[ 1\pm f(N,\beta )...\right] 
\ee

The first term leads to the classical regime addressed above; the
``correction'' term $f$ (negative/positive for fermions/bosons) in brackets
becomes:

\be
f(N,\beta )=\frac{Q(2\beta )}{Q^{2}(\beta )}=\frac{\rho \lambda _{T}}{\sqrt{2%
}}\left( \frac{1-e^{-b}}{1-e^{-2b}}\right) ^{2} 
\ee

Here $b=\beta \hbar \omega $. The factor preceding the expression in parentheses
is analogous to quantum statistical correction terms found in 2D and 3D expansions. We observe
that the classical approximation used earlier is appropriate when the
interparticle spacing exceeds $\lambda _T$ (the same constraint as that found in the
analogous 2D and 3D problems). The factor in parentheses in Eq. 20 is always less than one,
so it helps the statistical expansion to converge, especially at high T when
the factor becomes 1/4. The reason for this reduced
statistical correction is simply that the crowding in phase space, the
origin of effects of quantum statistics, is reduced by a spreading among the
many transverse states that are excited when $b<1$. Finally, we note that statistical corrections to the noninteracting classical
gas are relatively more important in 1D than in higher dimensions D. The reason
is
that these corrections appear as products of density and $\lambda_T^d$. At a given
T, therefore, the effect of statistics appears at a lower T in a system of
lower D (all other things being equal).

\section{Phonons in the high density groove phase}

The quasi-1D fluid within the groove represents a novel system in many ways.
One is that the low-lying excitations of the system are expected to be
phonons, even in the absence of a crystal (which is assumed in the usual
derivation of phonons in terms of an expansion in displacements from equilibrium). Such
phonons must exist as longitudinal, long wavelength excitations of this
system, derived from elasticity theory or quantum hydrodynamics (as in the
liquid helium case)\cite{dash}. Their speed for a nearly classical system can be derived from the classical equation of state, $mc^2=(dP/d\rho)_T$. Such data appear in Fig. 6. One observes a minimum in $c$ at density corresponding to the incipient condensation at T=0.

We analyze the behavior using the Debye model, applied to a fluid of density
$\rho$ and sound speed $c$. The Debye wave vector and frequency are $k_D=\pi \rho$ and $\omega_ D=c k_D$, respectively. The conventional 3D treatment is changed to accommodate
the 1D density of states, which is a constant below $\omega_ D$:

\be
N(\omega)= \frac{L}{\pi c}\, \Theta(\omega - \omega_D)
\ee

Here $\Theta (x)$ is the Heaviside unit step function. The resulting thermal energy per unit length is

\be
\frac{E}{L} = \frac{1}{\pi c\beta ^2\hbar} \int_0^{x_m} \,dx \, \frac{x}{e^x-1}
\ee

The upper limit to the integral arises from the Debye frequency
cutoff, $x_m=\beta \hbar \omega _D$. At high T, $x_m<<1$, this expression 
yields the 1D version of the law of Dulong and Petit,
$C \rightarrow N k_B$; there is energy $k_BT$ per atom in this limit. 
At low T, instead, the specific heat is linear in T:

\be
\frac{C}{N k_B} = \frac{\pi ^2 k_BT}{3\hbar \omega _D} 
\ee

A more realistic model would yield a different numerical coefficient but the
linear dependence of C on T is a robust prediction of the phonon model. The
prediction in the high T limit is also robust, within the harmonic expansion.

One may develop a concrete realization of this behavior from the usual
phonon theory. In its simplest form, we employ a model in which only nearest
neighbors interact. In this case, the ground state corresponds to a lattice
constant $a$ equal to the minimum in the pair potential. The conventional
theory yields a dispersion relation 

\be
\omega _{q}=2\omega _{0}\sin (\frac{qa}{2})\hspace{0.2in},\hspace{0.2in}\omega _{0}=\sqrt{\frac{k}{m}}
\ee

The force constant $k$ is just the second derivative of the pair potential,
evaluated at its minimum (if the adsorbate is not compressed). If we assume
a Lennard-Jones interaction $U(r)=4\varepsilon \left[(\sigma/r)^{12}-(\sigma/r)^{6}\right] $, then the
uncompressed force constant is $k_{0}=2^{8/3}(\frac{9\varepsilon }{\sigma
^{2}})$. This yields a speed of sound $c=a\omega _{0}=a\sqrt{k_{0}/m}$,
which depends in an explicit way on the specific system's parameters.

How important are the quantum corrections to the equation of state of the
system? One measure of this is the ratio of the system's ground state
zero-point energy (per atom) $E_{zp}$ to its potential energy ($\epsilon $); this ratio is zero for a classical system. From the phonon model, we may
evaluate $E_{zp}$ by summing $\hbar w_{q}/2$ over all of the phonon modes within
the Brillouin zone ( $|q|<\pi /a$ ). The result of this calculation is
 
\be
E_{zp}=\left( \frac{2}{\pi }\right) \hbar \omega _{0}
\ee

Hence, the ratio of interest is 

\be
\frac{E_{zp}}{\epsilon }=3 \left( \frac{2^{4/3}}{\pi ^2}\right) \Lambda^{\ast }
\ee

Here $\Lambda ^{\ast }=h/(\sigma \sqrt{m\epsilon })$ is the de Boer
quantum parameter and the numerical coefficient is about 0.8. Thus, the
ratio of $E_{zp}$ to $\epsilon $ is about 1.6 for H$_{2}$, 0.2 for CH$_{4}
$ and 0.05 for Xe, assuming ``typical'' values of the interaction
parameters. This number is indicative of the relative importance of quantum
effects at low T; the range of T over which quantum effects are relevant is
of order this ratio times $\epsilon $. The fully classical regime is thus
T exceeding 60 K, 35 K and 15 K for H$_{2}$, CH$_{4}$ and Xe, respectively. 

The preceding discussion pertains to the longitudinal phonons in the groove.
There are, in addition, transverse phonons associated with motion
perpendicular to the groove axis. We show here how these become important at
high density, indicative of an incipient instability of the 1D state. The transverse modes (two polarizations for each wave vector) can be derived in the usual way by taking account of
coupled motion of the adsorbate, as affected by the presence of the
potential confining the particles within the groove ($\alpha r^{2}/2$).
Consider a mode polarized in the $x$ direction and assume small amplitude
displacements \{$x_{i}$\} so that the interparticle force can be expanded
about the equilibrium spacing $a$; for a given pair, the potential energy is
thus shifted by

\be
V(\Delta x)-V_{0}=V^{\prime }[\frac{\Delta x^{2}}{2a}]=f\frac{\Delta x^{2}}{2}
\ee

Here, $V_{0}$ is the equilibrium spacing potential, prime means
derivative with respect to the interparticle spacing, evaluated at spacing $a
$, and $\Delta x$ is the difference in the $x_{i}$ values of adjacent
particles. The quantity $f$ is an effective force constant associated with
this coupling; $f<0$ in a highly compressed phase. Assuming only nearest
neighbor interactions, the equation of motion for a mode of wave vector $q$
can be solved, yielding the transverse spectrum 

\be
\omega _{t}^{2}(q)=\omega _{1}^{2}+(4\frac{f}{m})\sin^{2}(\frac{qa}{2})\hspace{0.2in},\hspace{0.2in}\omega _{1}^{2}=\frac{\alpha }{m}
\ee

At $q=0$, the frequency is just that ($\omega _{1}$) of single particles in
the external potential. However, the finite $q$ frequency is lower if the
adsorbate is compressed. At the zone boundary, $\omega $ is a minimum in
this case. There arises, therefore, an instability when

\be
f<-m\omega _{1}^{2}/4=-\alpha /4
\ee

This condition is equivalent to the energy minimum condition, which favors a
periodic structure if the density becomes high, so that the repulsive forces
become too large. The preceding equation represents an instability
criterion, but one expects there to be a lower density regime of dynamical
stability but energetic metastability. We have explored this problem in
recent work by comparing the energies of alternative structures on the
external surface of the bundle \cite{exsurf2}. We concluded that the transition in question
occurs from the groove to the three-stripe phase. An alternative
possibility \cite{zig}, a two-stripe phase (two parallel chains of atoms), was less
favored in the Lennard-Jones interaction case we have considered. 

We have made a quantitative comparison with the phonon instability 
scenario in one case, CH$_{4}$ in a groove. In
that case, the transition to the three stripe phase occurs at 10 \%
compression in lattice constant, according to the T=0 calculations. The instability condition above corresponds
to a further compression of the lattice constant by 20 \%. Hence, the 1D phase becomes metastable (relative to the three-stripe
phase) at considerably lower density, preempting the instability-driven
transition.

In the limit of small compression, we may neglect the term in Eq. 28 proportional to $f$. In that case, the transverse motion reduces to that of independent particles. The resulting effect on the thermal properties of the adsorbate coincides with that of single particles, i.e. the low density limit, discussed in the Appendix.

\section{Coverage dependent adsorption: grand canonical Monte Carlo simulations}

As in our previous studies \cite{exsurf1,exsurf2}, we use grand canonical Monte Carlo
simulations to investigate the adsorption behavior of CH$_{4}$ on
the external surface of a bundle. The potential energy 
experienced by the adsorbate particles in that region is modeled by summing the
contributions from two adjacent nanotubes, by adding Lennard-Jones two body interactions (with distance and energy parameters $\sigma _{gc} = 3.56$ \AA, $\epsilon _{gc}=67.2$ K) between the adsorbate particle and the nanotube's carbon atoms. The
dimensions of the simulation cell are set to be 17 \AA$\,$ in the $x$ direction
(which corresponds to the center to center distance between the tubes), 10 $\sigma _{gg}$ along the $z$ direction (tube's axis direction) and 40 \AA$\,$ for
the height of the box along the $y$ direction. The details of our model
assumptions and the simulation method can be found in Ref. 17.

Figure 7 shows the resulting CH$_{4}$ adsorption isotherms at
various temperatures. Since the length of the cell in the $z$ direction 
is 10 $\sigma _{gg}$, the saturation number of atoms in a single line 
(closed packed) is $\langle N \rangle=9$. The phase behavior evolves 
with $\mu$ in the same manner
 observed for other gases \cite{exsurf1,exsurf2}, starting with a line 
of atoms in the groove at very low pressure. At low T, a step (not 
noticeable for the temperatures shown in the figure) appears to a 
so-called ``3-stripe'' phase, consisting of two additional lines of 
atoms parallel to that in the groove. At higher pressure a monolayer 
($\langle N \rangle \approx$ 45, 5 lines) is formed over the external 
surface. Once it completes, there appears some evidence of a
transition to a second layer groove phase ($\langle N \rangle \approx$
54), i.e. a single line of atoms formed above the monolayer phase, in
the new groove region. This phase was observed in our previous
simulations in the case of Ne, Ar and Kr and also experimentally in
recent work of Migone's group \cite{aldosg,aldosg2}. Subsequent 
transitions to bilayer and three-layer phases are observed as the 
pressure increases. 

In figure 8 we display the evolution of the density projected onto
the transverse $xy$ plane at $T=90$ K as the coverage increases, starting at
the monolayer phase. We observe the formation of the second layer groove, a
dilute second layer 3-stripe phase and the completion of the 
second layer. The variation of potential along the second layer surface
 is still large enough to gives rise to the
appearance of a third-layer groove. After that, the adsorbate surface has
considerably flattened at this point, causing the subsequent film to be much less structured. 

We consider now the formation of the second layer groove phase for 
three different gases of increasing size: Ne, CH$_4$ and Xe. Figure 
9 shows its respective isotherms for high coverage. The arrows point 
the appearance of the second layer groove. Table II displays the values of the pressure from the isotherms at which this phase occurs in comparison with the experimental values \cite{aldosg2}, indicating a very good agreement. 
To identify this
phase we also show the corresponding density contours (Fig. 10) and the
compresibility $dN/d\mu$ (Fig.  11) calculated from the number
fluctuation in the simulations:

\be \frac{dN}{d\mu}=\frac{\langle (\Delta N)^2 \rangle}{k_B T} \ee

Here $\langle (\Delta N)^2 \rangle$ is the variance in the number of
particles in the simulation (at fixed $\mu$ and T). For the smallest 
atom, Ne, there occurs a peak near
$\langle N \rangle \approx 80$, which corresponds to the
(best-defined) second layer groove phase.  Then, transitions in the
second layer (three-stripe phase and completion of the second layer)
cause the appearance of a second broader peak at $\langle N \rangle
\approx 100-110$, but they cannot be distinguished individually. For
CH$_4$, there is a smaller peak near $\langle N \rangle \approx 60$
(second layer groove) but it merges with the peak corresponding to the
second layer transition ($\langle N \rangle \approx 70$). Something
similar happens in the case of Xe. The peak starts with the second layer groove transition ($\langle N \rangle \approx 55$) but it immediatly continues with the bilayer transition ($\langle N \rangle \approx 80$). 

Another quantity which should show the phase transitions as the coverage 
increases is the isosteric heat. 
Figure 12.(a) shows the isosteric heat as a function of linear coverage
computed from adsorption isotherms of the different gases. The region
$\rho \sigma_{gg} < 1$ is the groove-filling region. Once the groove is
filled, the isosteric heat decreases abruptly due to the high energy
difference between this site and the surface site. The following
decrease is observed at the monolayer completion but before that a small increase strongly suggests the presence of the second layer groove. 

In Tables III and IV, we compare the isosteric heat values corresponding to the first groove phase and the monolayer phase with the available experimental results from two different groups. We observe that the agreement is quite good.

In Fig. 12.(b) we compare the whole coverage dependence of the isosteric heat for Ar derived from our simulations with the experimental results \cite{oscar1}. We observe that the general trend is qualitatively similar.  

Fig 13 shows the specific heat as a function of coverage for Kr and Ar
calculated from the simulations data.  In both cases, a notable
increase is observed near the completion of the groove and a smaller
one is present at the end of the monolayer completion, very possibly due to
the presence of the second groove phase. The low density value ($N<5$) agrees reasonably well with the low density limit calculations of Section II for those gases and temperatures.

\section{Summary and conclusions}

Our calculations have yielded classical and quantum behavior 
of diverse
gases adsorbed on the external surface of a nanotubes bundle. 
The present results for thermal properties are to be supplemented, 
in general,
by contributions from gases adsorbed in other sites, i.e. the 
interstitial
channel and endohedral positions, if these are accessible to the
adsorbate. The degree to which this is the case appears to be very
sensitive to sample preparation and purification technique. 
Evidence in
this paper's tables provides some tentative support for the belief that a
significant fraction of these samples' area is ordered and clean. Yet Fig.12(b) shows a qualitative discrepancy, presumably attributable to heterogeneity.

We have explored the problem of dimensional crossover by studying
adsorption along two distinct thermodynamic paths: constant T 
(variable N) and variable T (very low N). At very low coverage, the
effective dimensionality increases progressively with T because the
adsorbed molecules migrate over a T-dependent phase space, 
beginning (at low T) with the groove and ending with monolayer 
and even 3D regimes at higher T. Qualitatively similar evolution 
was investigated some years ago
in the case of He isotopes on graphite \cite{mil81}. In
that case, the T dependence of the dimensionality (calculated and measured) 
reflects the energy dependence of the wave functions' spatial 
localization. The more conventional method of studying dimensional 
crossover is to assess the variation with coverage of film 
structure and thermal
properties. This is the route followed in adsorption
isotherm measurements. Evidently, one can (in principle) explore 
the N-T plane along any path. An interesting question that we have 
not explored is
how the effective dimensionality of the film varies at higher T 
or N than is reported here, yielding a more complete 
characterization of the effective dimensionality. We note that 
specific heat and isotherm experiments provide complementary 
information, so that both experiments
are worth carrying out \cite{good}. To our knowledge,
no specific heat measurements have yet been undertaken for gas 
adsorption on nanotubes. This situation is probably temporary, 
because the very high
specific area found in nanotubes samples should yield very high 
total heat capacities, with a relatively small background correction at low T. This
argument suggests that measurement and interpretation of C(N,T) 
data are
likely to be fruitful and convenient.

Our computational methods used in this paper are
relatively straightforward, i.e. mostly extensions of those used in our previous
simulation studies. Hence, we have found few surprises in the 
results. One
of the most intriguing findings is that the second layer groove 
phase is
present in the isotherms (dramatically so in the compressibility) for all
of the systems studied, consistent with experiments of the Migone group.
Equally encouraging is the agreement reported in the previous section
between these calculations and experimentally observed thermodynamic
quantities. Such consistency is initially surprising, in view of the
simplified potential models. One concludes that the interaction strengths
are adequately transferable from the graphite adsorption problem. 
Such
behavior is not consistent with some model calculations in which 
either
curvature-induced distortion of the physisorption potential or sensitivity
of the potential to the nanotubes' conductivity is present.

\begin{acknowledgments}
We wish to acknowledge Mary Jo Bojan, Bill Steele, Silvina Gatica, Milen Kostov,
Aldo Migone, J.G. Dash, Karl Johnson, Oscar Vilches and Michel Bienfait for 
stimulating
discussions. We are grateful to the Petroleum Research Foundation of
the American Chemical Society, the Army Research Office and Fundaci\'on Antorchas for their support. 
\end{acknowledgments}

\bigskip

\appendix*
\section{1D Motion}

\bigskip

In the limiting case of motion in 1D with a quadratic transverse
confining potential

\bigskip

\be V(r)=V_{0}+\alpha \frac{r^{2}}{2} \ee

\noindent where $r^{2}=x^{2}+y^{2}$ and the $z$ axis is the usual
cylindrical symmetry axis, the classical configuration integral
satisfies

\bigskip \bigskip \be Z(\beta )=\frac{2\pi L}{\beta \alpha }e^{-\beta
V_{0}} \ee

\bigskip

From Eq.8 and 9, classical values of the thermal variables in this
quasi-1D limit are

\bigskip 
\begin{eqnarray*}
q_{cl} &=&-V_{0} \\ \frac{C_{cl}}{Nk_{B}} &=&\frac{5}{2} \\ \mu _{cl}
&=&V_{0}+\frac{1}{\beta }\ln \left( \frac{\rho \lambda _T}{\pi }
\frac{\lambda _T^2}{\langle r^2 \rangle}\right)
\end{eqnarray*}

\bigskip
\noindent where $\langle r^2 \rangle=2/(\alpha \beta)$ is the mean
square particle displacement perpendicular to the axis.  The analogous
quasi-1D quantum values are given by assuming that the temperature is
sufficiently low that only the lowest transverse vibration is present,
for which the energy is 
$\varepsilon _{0}=V_{0}+\hbar \omega , \,\omega =\sqrt{\alpha /m}.$

\bigskip In this case, $g(E)=G_{1}(E-\varepsilon
_{0})\Theta(E-\varepsilon _{0})$
, where $G_{1}(x)= \frac{L}{\hbar \pi} \sqrt{\frac{m}{2x}}$. 
This is the 1D density of states for a
hypothetical system with no transverse degrees of freedom. Then, 
the single particle partition function is

\be 
Q(\beta )=\frac{L}{\lambda }e^{-\beta \varepsilon _{0}} 
\ee

A generalization of this expression to include all transverse states,
within the harmonic approximation, leads to correction to $Q$ by a
factor, $\left( 1-e^{-\beta \hbar \omega }\right) ^{-2}$.  Without
these factors,

\bigskip 
\begin{eqnarray*}
\frac{C}{Nk_BT} &=&\frac{1}{2} \\ q_{st} &=&\frac{2}{\beta
}-\varepsilon _{0} \\ \mu &=&\varepsilon _{0}+\frac{1}{\beta }\ln
(\rho \lambda )
\end{eqnarray*}

\bigskip

Note that the quantum isosteric heat exceeds the classical value by an
amount $\delta q_{st}$ given, in the present approximation, 
by $\delta
q_{st}=q_{st}-q_{cl}=2/\beta -\hbar \omega$.  This is, as expected, the
difference in energy associated with the quantized harmonic motion in
x and y directions. We discuss in Section III the more general case,
when other transverse degrees of freedom are excited.  The present
result applies to the low T regime, where $\hbar \omega >>2/\beta $. 

\newpage

\begin{table}
\begin{tabular}{|c|c|c|c|} \hline \hline
             &\,\,Theory\,\,&\,\,Exp.1\,\,&\,\,Exp.2\,\, \\ \hline
\,\,H$_2$\,\,&    1.43      &1.5      &     1.5         \\ \hline
\,\,D$_2$\,\,&    1.45      &   -      &     1.8    \\ \hline
\,\,Ne   \,\,&    1.51      &1.73    &      -     \\ \hline
\,\,CH$_4$\,\,&   1.43      &1.76     &     1.34  \\ \hline
\,\,Xe    \,\,&   1.41      &1.74     &     1.37  \\ \hline \hline
\end{tabular}
\caption{
Ratio of binding energy of a molecule in the groove to that on graphite. The values from theory correspond to the limit of zero coverage (single particle). The experimental values are obtained from isosteric heat values through the relation $\varepsilon_0=-q_{st}+2k_BT$, assuming that the adsorbed phase behaves as a 1D system in the temperature and density range explored. Exp.1 values are from Ref. 9 and Exp.2 values from Ref. 13 (H$_2$) and 14 (CH$_4$ and Xe).}
\end{table}

\begin{table}
\begin{tabular}{|c|c|c|c|} \hline\hline
     \, &\,\,T(K)\,\,&\,\,GCMC\,\,&\,Experiment\,\\ \hline 
\,Ne \, &
25 & -1.9 & -  \\ \hline \,CH$_4$\, & 70 & -3.4 & -3.2 \\ \hline
\,Xe \, & 112 & -2.4 & -2.6 \\ \hline\hline
\end{tabular}
\caption{Common logarithm of the pressure (atm) at which the second
groove phase appears. The experimental values are from Ref. 11.} 
\end{table}

\begin{table}[h]
\begin{tabular}{|c|c|c|c|c|c|} \hline\hline
     \, &\,\,T\,(K)\,\,&\,\,$\epsilon_{gg}$\,(K)\,\,&\,\,GCMC\,\,&\,Exp.1\,&\,Exp.2\,\\ \hline 
\,Ar\,    & 90  & 120  & 13.9 & -    & 15.2 \\ \hline 
\,CH$_4$\,& 90  & 161  & 12.4 & 10.3 & 13.5 \\ \hline 
\,Xe\,    & 110 & 221  & 12.4 & 12.5 & -    \\ \hline\hline
\end{tabular}
\caption{Isosteric heat $q_{st}/\epsilon_{gg}$ at a typical first groove
coverage. Exp.1 values are from Ref. 8 (CH$_4$) and 10 (Xe). Exp.2
values are taken from Ref. 13 (Ar) and 14 (CH$_4$).} 
\end{table}

\newpage

\begin{table}[h]
\begin{tabular}{|c|c|c|c|c|c|} \hline\hline
     \, &\,\,T\,(K)\,\,&\,\,$\epsilon_{gg}$\,(K)\,\,&\,\,GCMC\,\,&\,Exp.1\,&\,Exp.2\,\\ \hline 
\,Ar\,    & 90  & 120 & 10.0 & -   & 10.1 \\ \hline 
\,CH$_4$\,& 90  & 161 & 9.3  & 9.0 & 8.3  \\ \hline 
\,Xe\,    & 110 & 221 & 8.6  & 9.9 & 8.5  \\ \hline\hline
\end{tabular}
\caption{Isosteric heat $q_{st}/\epsilon_{gg}$ at a typical 
monolayer
coverage. Exp.1 values are from Ref. 8 (CH$_4$) and 10 (Xe). Exp.2
values are taken from Ref. 13 (Ar) and 14 (CH$_4$ and Xe). }
\end{table}

\newpage

FIGURE CAPTIONS

\vspace{1cm}

FIG. 1: Isopotential contours for a CH$_4$ molecule on the
external surface of a nanotube bundle. The contour nearest the groove ($x=y=0$) corresponds to $V=-2000$ K.

FIG. 2: The low density specific heat, in units of Boltzmann's
constant, for CH$_4$. Full curve is the classical 
specific heat
obtained from Eq.9 and the potential energy shown in Fig.1. Thin
curve is the classical specific heat derived from the model 
crossover potential. Dashed
curve is the quantum specific heat from Eqs.5 and 15.

FIG. 3: Effective potential energy V$_{eff}(x)$ (dashed curve)
obtained by solving the Schrodinger equation for the CH$_4$ motion
perpendicular to the surface at each position $x$ on the 
surface. The full curve is a fit to this potential using the functional form of Eq. 13.

FIG. 4: Isosteric heat and quantum corrections for CH$_4$ in the three cases plotted in Fig.2.

FIG. 5: (a) Same as Fig. 2 for Ne. (b) Same as Fig. 4 for Ne.

FIG. 6: Speed of sound for CH$_4$ in the groove as a function of reduced density at
different values of reduced temperature T$^*$=T/$\epsilon _{gg}$. 

FIG. 7: CH$_4$ adsorption isotherms at several temperatures. Saturation capacity of the groove occurs at $\langle N \rangle = 9$. The monolayer coverage is $\langle N \rangle \approx 45$.

FIG. 8: Evolution of the CH$_4$ density with P, at $T=90$ K. From top to bottom, in the left column: monolayer phase (P = 0.85 10$^{-2}$ atm); second layer groove phase (P = 0.17 10$^{-1}$ atm); second layer 3-stripe phase (P = 0.20 10$^{-1}$ atm). In the right column: bilayer phase (P = 0.30 10$^{-1}$ atm); third layer groove phase (P = 0.37 10$^{-1}$ atm); multilayer phase (P = 0.52 10$^{-1}$ atm). 

FIG. 9: Different gases' isotherms showing the formation of the second layer groove phase, occurring at points indicated by arrows.

FIG. 10: Density contours projected onto the x-y plane showing the second layer groove phase for different gases. From top to bottom: Ne (T = 28 K, P = 0.06 atm), CH$_4$ (T = 90 K, P = 0.017 atm), and Xe (T = 112 K, P = 0.005 atm).

FIG. 11: Compressibility as a function of coverage for different 
gases and coverages beyond the first layer.  The curves are 
guides for the eye.

FIG. 12: (a) Reduced isosteric heat as a function of coverage for Ne, CH$_4$ and Xe at T$^* \approx 0.75$. $\epsilon _{gg}=$ 35.6 K (Ne), 161 K (CH$_4$), 221 K (Xe).(b) Same as (a) but for Ar ($\epsilon _{gg}=$ 120 K). The dots represent experimental results from Ref. 13.

FIG. 13: Heat capacity derived from classical simulations as a function of coverage for Ar at T=67.5 K (dashed curve) and Kr at T=71.25 K (full curve).


\begin{references}

\bibitem{rmp} M.M. Calbi, M.W.  Cole, S.M.  Gatica, M.J. Bojan and
G. Stan, Rev. Mod. Phys. {\bf 73}, 857 (2001). 

\bibitem{will} K.A.  Williams and P.C. Eklund, Chem.  Phys. Lett. {\bf
320}, 352 (2000). 

\bibitem{wang} Q. Wang, S. R. Challa, D. S. Sholl, and J. K. Johnson,
Phys. Rev. Lett. {\bf 82}, 956 (1999). 

\bibitem{boro} J.  Boronat, M.C. Gordillo, and J. Casulleras, J. Low
Temp.  Phys.  {\bf 126}, 199 (2002); M.C.  Gordillo, J.  Boronat,
J. Casulleras, Phys. Rev.  B {\bf 65}, 014503 (2002);
M.C.  Gordillo, J.  Boronat, J.  Casulleras, J.  Low Temp.  Phys. {\bf
121}, 543 (2000); M.C.  Gordillo, J.  Boronat, J.  Casulleras,
Phys. Rev. Lett. {\bf 85}, 2348 (2000). 

\bibitem{bon} M.  Boninsegni, S.Y.  Lee, V.H.  Crespi,
Phys. Rev. Lett. {\bf 86}, 3360 (2001). 

\bibitem{dil} M.M.  Calbi, F.  Toigo, and M.W.  Cole,
Phys. Rev. Lett. {\bf 86}, 5062 (2001). 

\bibitem{bob} Y.H. Kahng, R.B. Hallock, E. Dujardin, and T.W. Ebbesen,
J. Low Temp. Phys.  {\bf 126}, 223 (2002); W. Teizer, R.B. Hallock,
E. Dujardin, and T.W. Ebbesen, Phys. Rev. Lett. {\bf 82}, 5305 (1999);
erratum {\bf 84}, 1844 (2000). 

\bibitem{aldo} S. Talapatra, A.D.  Migone, Phys.  Rev.  B {\bf 65}, 045416 
(2002). 

\bibitem{aldo2} A.J. Zambano , S. Talapatra, A.D. Migone, Phys. Rev. B
{\bf 64}, 075415 (2001); S. Talapatra, A.Z. Zambano,
S.E. Weber, and A.D. Migone, Phys.  Rev. Lett. {\bf 85}, 138 (2000);
S.E. Weber, S. Talapatra, C. Journet, A.  Zambano, and A.D. Migone,
Phys. Rev. B {\bf 61}, 13150 (2000). 

\bibitem{aldosg} S. Talapatra, A.D. Migone, Phys. Rev. Lett. {\bf 87},
206106 (2001). 

\bibitem{aldosg2} S. Talapatra, A.D.  Migone, {\it Higher Coverage Gas
Adsorption on the Surface of Carbon Nanotubes: Evidence for a New
Phase in the Second Layer}, unpublished. 

\bibitem{yates} A.  Kuznetsova, J.T.  Yates, V.V.  Simonyan,
J.K. Johnson, C.B. Huffman, R.E. Smalley, J.  Chem. Phys.  {\bf 115}
6691 (2001); V.V. Simonyan, J.K. Johnson, A.  Kuznetsova, J.T. Yates,
J.  Chem. Phys.  {\bf 114}, 4180 (2001); A. Kuznetsova, J.T. Yates,
J. Liu, R.E. Smalley, J. Chem. Phys. {\bf 112}, 9590 (2000). 

\bibitem{oscar1} T.  Wilson, A.  Tyburski, M.R. DePies, O.E. Vilches,
D. Becquet, and M. Bienfait, J. Low Temp. Phys. {\bf 126}, 403 (2002). 

\bibitem{oscar} M.  Muris, N.  Dupont-Pavlovsky, M.  Bienfait, and
P. Zeppenfeld, Surf. Sci. {\bf 492}, 67 (2001); M.  Muris, N. Dufau,
M. Bienfait, N. Dupont-Pavlovsky, Y. Grillet, J.P. Palmari, Langmuir
{\bf 16}, 7019 (2000). 

\bibitem{kaneko} K.  Murata, K.  Kaneko, W.A.  Steele, F.  Kokai,
K. Takahashi, D. Kasuya, M. Yudasaka, S. Iijima, Nano Letters {\bf 1},
197 (2001); T. Ohba, K. Murata, K. Kaneko, W.A. Steele, F. Kokai,
K. Takahashi, D. Kasuya, M. Yudasaka, S. Iijima, Nano Letters {\bf 1},
371 (2001). 

\bibitem{kan&bill} H. Tanaka, M.  El-Merraoui, W.A. Steele, K. Kaneko,
Chem. Phys. Lett. {\bf 352}, 334 (2002). 

\bibitem{exsurf1} S.M. Gatica, M.J.  Bojan, G.  Stan, and M.W. Cole,
J. Chem. Phys. {\bf 114}, 3765 (2001). 

\bibitem{exsurf2} M.M. Calbi, S.M.  Gatica, M.J. Bojan, and M.W. Cole,
J. Chem. Phys. {\bf 115}, 9975 (2001). 

\bibitem{george} G. Stan and M.W.  Cole, Surf.  Sci. {\bf 395}, 280
(1998). 

\bibitem{milen} Recent work suggests that the long range gas-gas
interaction is significantly screened by carbon atoms.  See
M.K.  Kostov, M.W.  Cole, J.C.  Lewis, P.  Diep, J.K.  Johnson,
Chem. Phys. Lett. {\bf 332}, 26 (2000). 

\bibitem{bruch} L.W.  Bruch, M.W. Cole and E. Zaremba, {\it Physical
Adsorption: Forces and Phenomena} (Oxford University Press, 1997), page 66. 

\bibitem{henry} The conventional quantity used in this definition 
is the thermodynamic excess coverage, rather than N. In the present 
case of very attractive potentials, the diference between these 
quantities is small.

\bibitem{spin} Spin and other internal degrees of freedom, if any, 
change computed quantities by an additive constant, at most, 
unless there is coupling to an external field. M.K. Kostov et 
al (J. Chem. Phys. 116, 1720, 2002), for
example, found a large effect of rotational hindrance of hydrogen
molecules inside grooves and interstices of nanotube bundles. Experimental support for this was found by D.G. Narehood, P.E. Sokol, P.C. Eklund, M.K. Kostov and M.W. Cole, Phys. Rev. B in press.

\bibitem{qanah} The quantum expression is $(a_1+6a_2)\langle x^2
\rangle ^2$ where $\langle x^2\rangle=\frac{\hbar\omega}
{2\alpha} \coth(\frac{\beta\hbar\omega}{2})$.

\bibitem{adiab} M.W.  Cole and F. Toigo, Phys. Rev. B {\bf 23}, 3914
(1981). 

\bibitem{qmbook} I.I. Gol'dman and V.D. Krivchenkov, {\it Problems in
Quantum Mechanics} (Dover Publications, Inc., New York, 1993). 

\bibitem{dash} J.G. Dash, Rev. Mod. Phys. {\bf 71}, 1737 (1999).

\bibitem{zig} P. Zeppenfeld, M. Muris, M. Bienfait, N. Dupont-Pavlovsky, and M. Johnson (to be published).

\bibitem{mil81} M.W. Cole, D.R. Frankl, and D.L. Goodstein, Rev. Mod. Phys. 53, 199 (1981). 

\bibitem{good} R.L. Elgin and D.L. Goodstein, Phys. Rev. A {\bf 9}, 2657 (1974).

\end{references}
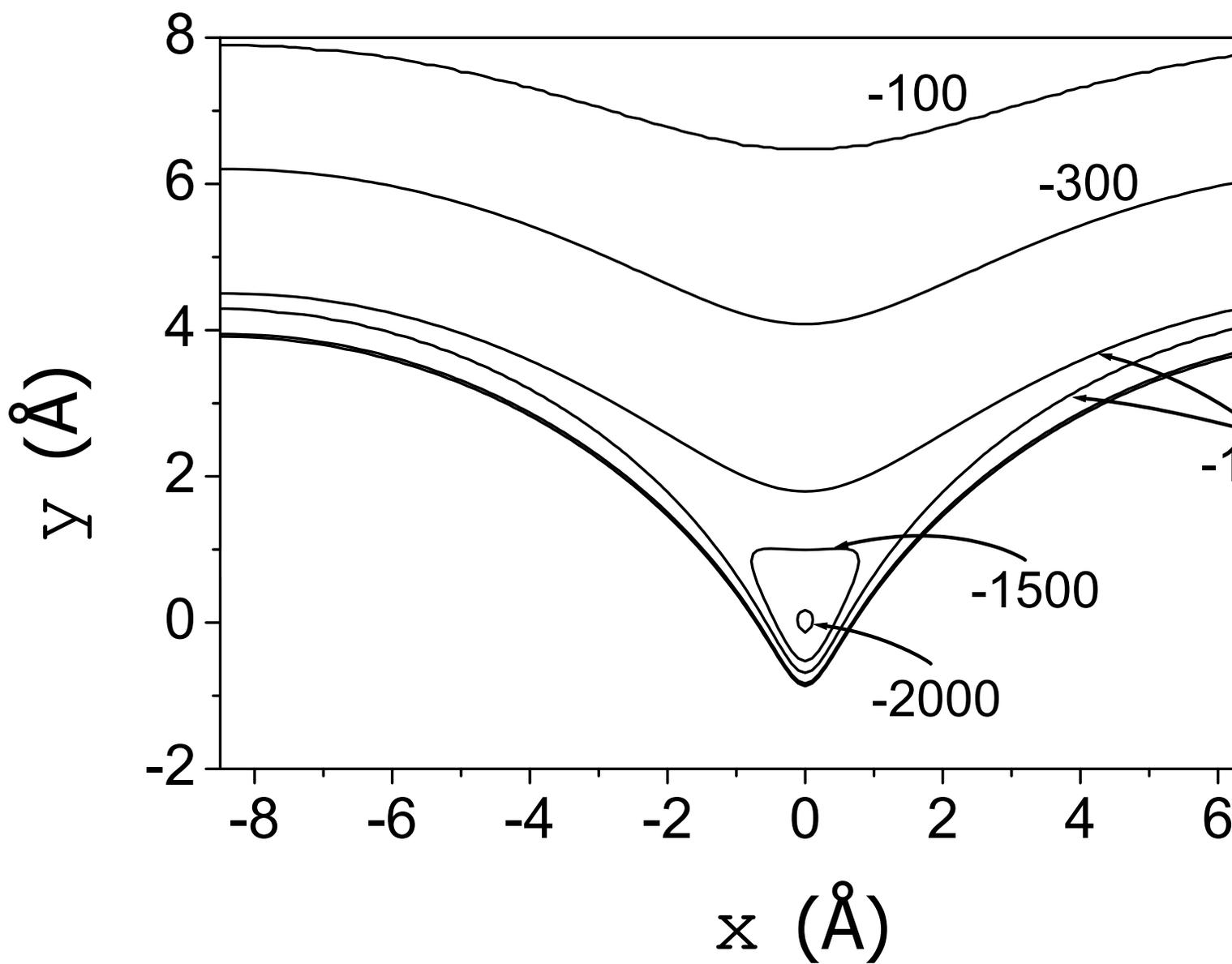
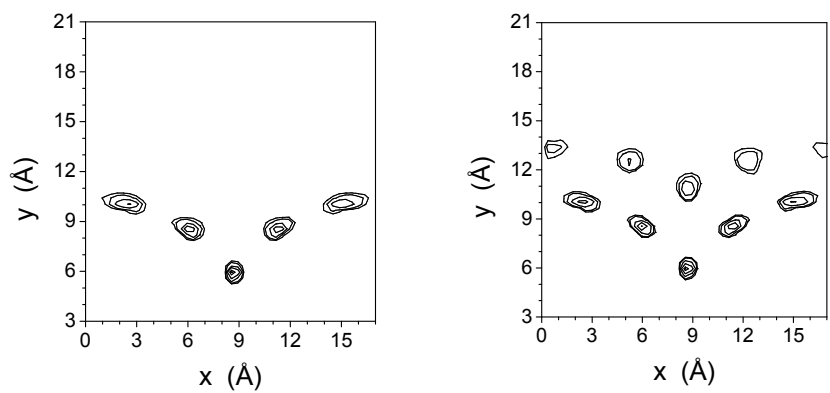
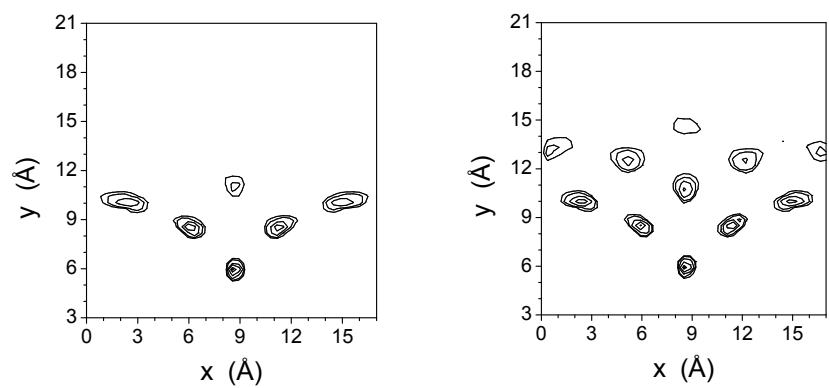
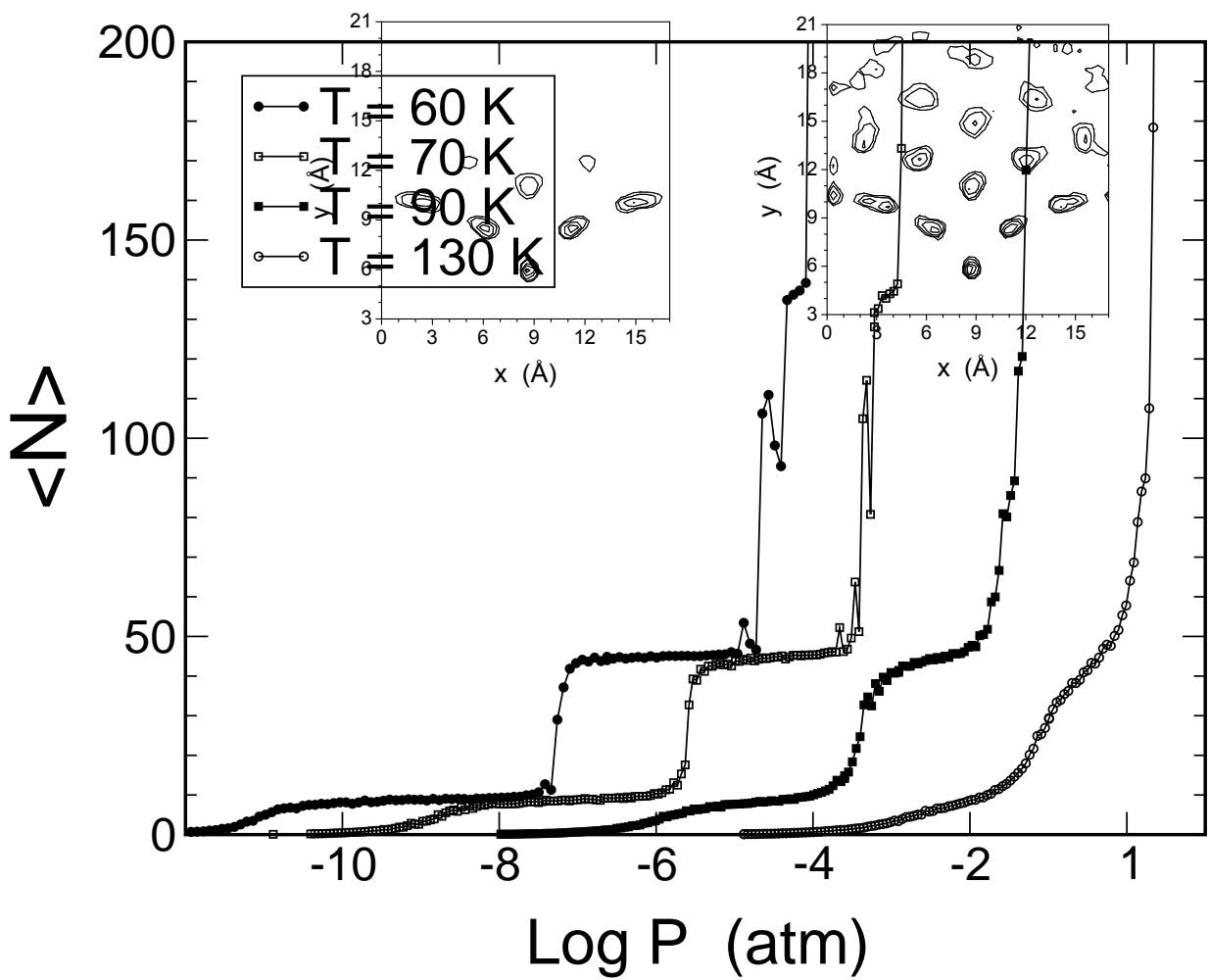
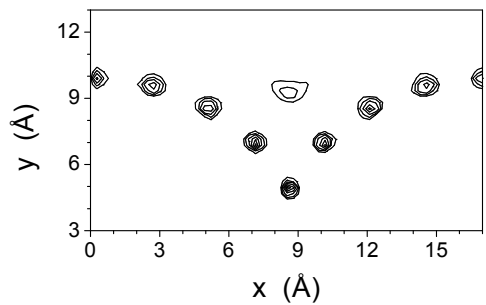
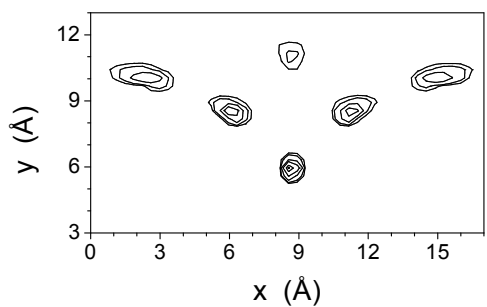
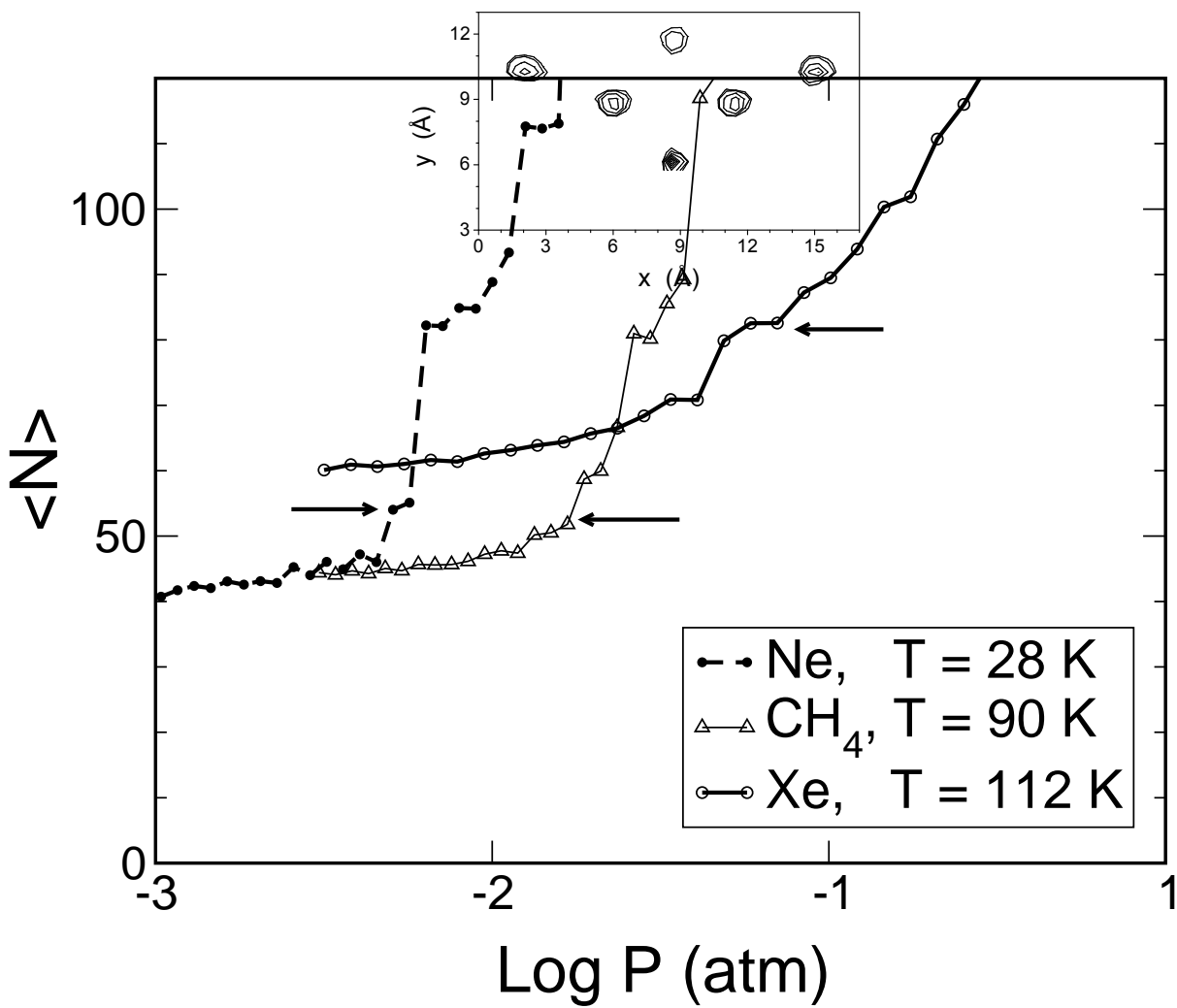
\end{document}